\documentclass[aps, reprint]{revtex4-1}

\usepackage[left=1.7cm,top=1.6cm,right=1.7cm,bottom=2.9cm]{geometry}
\usepackage[latin1]{inputenc}
\usepackage{amsfonts}
\usepackage{amssymb}
\usepackage{amsmath}
\usepackage{calc}
\usepackage{graphicx}
\usepackage{epsfig}
\usepackage{bm}
\usepackage{ulem}
\usepackage{setspace}
\usepackage{enumerate}
\usepackage{wasysym}
\usepackage{natbib}

\begin{document}

\title{\large \bf Purification of Single-Photon Entanglement}

\author{\normalsize D. Salart,\textsuperscript{1} O. Landry,\textsuperscript{1} N. Sangouard,\textsuperscript{1} N. Gisin,\textsuperscript{1} H. Herrmann,\textsuperscript{3} B. Sanguinetti,\textsuperscript{1} C. Simon,\textsuperscript{1,2} W. Sohler,\textsuperscript{3}\\ R. T. Thew,\textsuperscript{1} A. Thomas,\textsuperscript{3} and H. Zbinden\textsuperscript{1}}

\address{\it \small \textsuperscript{1}Group of Applied Physics, University of Geneva, 20, Rue de l'Ecole de M\'edecine, CH-1211 Geneva 4, Switzerland\\ \textsuperscript{2}Present address: Institute for Quantum Information Science and Department of Physics and Astronomy, University of Calgary, 2500 University Drive NW, Calgary T2N 1N4, Alberta, Canada\\ \textsuperscript{3}Universit\"{a}t Paderborn, Fakult\"{a}t für Naturwissenschaften, Department Physik, Warburger Straße 100 33098 Paderborn, Germany}

\date{\small \today}

\begin{@twocolumnfalse}
\begin{abstract}
{\small \parbox[t]{14cm}{\hspace{2mm} Single-photon entanglement is a simple form of entanglement that exists between two spatial modes sharing a single photon. Despite its elementary form, it provides a resource as useful as polarization-entangled photons and it can be used for quantum teleportation and entanglement swapping operations. Here, we report the first experiment where single-photon entanglement is purified with a simple linear-optics based protocol. Besides its conceptual interest, this result might find applications in long distance quantum communication based on quantum repeaters.}}
\end{abstract}
\end{@twocolumnfalse}

\maketitle

Entanglement purification provides a fascinating conceptual viewpoint to gain insight into the properties of entanglement. It can be used for the quantification of entanglement in bipartite systems \cite{Horodecki}. It may also be useful in practical applications, e.g. in the context of long distance quantum communication where the direct transmission of photons through an optical fiber is limited by losses and the no-cloning theorem. This can be overcome using quantum repeaters \cite{Briegel98}, which require the creation of entanglement over short links, the storage of entangled states within these links, and entanglement swapping operations to distribute entangled states over longer distances. In practice, these operations introduce errors, limiting the number of links that can be used. While the most immediate goal of outperforming the direct transmission may not need purification, the entanglement distribution within future quantum networks requires a larger number of links, necessitating several purification operations \cite{Sangouard09}. 

Initial proposals by Bennett {\it et al.} \cite{Bennett96} and Deutsch {\it et al.} \cite{Deutsch96} for entanglement purification were expressed in terms of quantum gates. For practical applications e.g. in the frame of quantum repeaters, it is important to keep implementations as simple as possible. For example, the protocol presented in Ref. \cite{Pan01} and implemented as reported in Ref. \cite{Pan03} requires linear optical elements only, and can easily be integrated into quantum repeater architectures. However, this last proposal is suited to the purification of polarization-entangled pairs of photons whereas many attractive quantum repeater protocols \cite{Duan01,Simon07,Sangouard07} and related experiments \cite{Chou05,Choi} use single-photon entanglement, i.e. entanglement of the form $|1\rangle_A|0\rangle_B+|0\rangle_A |1\rangle_B$, where two modes $A$ and $B$ share a single photon. First, these repeaters are rather simple: they require significantly fewer resources than other protocols and are thus less sensitive to memory and photon detector inefficiencies \cite{Sangouard09}. Furthermore, these quantum repeaters are efficient since they offer high entanglement distribution rates when combined with temporal multiplexing \cite{Simon07}. The main drawback of protocols based on single-photon detections is that, unlike protocols based on two-photon detections, they are interferometrically sensitive to path length fluctuations \cite{Minar07} that are at the origin of phase errors. Purification of single-photon entanglement is thus particularly important in this context. Here, we report the first experimental implementation of a protocol for phase-error purification of single-photon entanglement based on linear optics.

\begin{figure}
\includegraphics[width=0.8\linewidth]{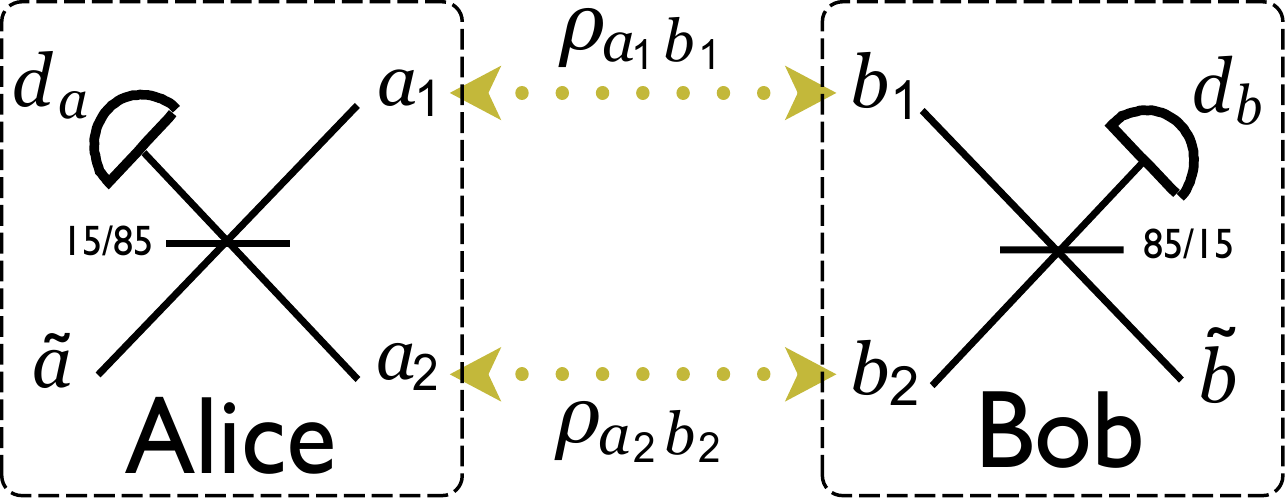}
\caption{\label{fig1} (color online). Scheme for entanglement purification of single-photon entanglement. Alice and Bob share two entangled
single-photon states $\rho_{a_1b_1},$ $\rho_{a_2b_2}$ (of the form given in Equation (\ref{mixedstate})) with
fidelity $F_1$ and $F_2$, respectively. While Alice combines
the modes $a_1$ and $a_2$ on a 15/85 beam splitter, Bob 
couples the modes $b_1$ and $b_2$ on a 85/15 beam splitter.
The detection of one photon in either $d_a$ or $d_b$ projects the
modes $\tilde{a}$ and $\tilde{b}$ into a single-photon
entangled state $\rho_{\tilde{a}\tilde{b}}$ with higher fidelity $\tilde{F}$.}
\end{figure}

The principle of purification for phase errors (see Ref. \cite{Sangouard08pur} for details) can be illustrated as follows. Alice and Bob, two protagonists located at remote locations $A$ and $B$ respectively, wish to share a maximally entangled state $\psi_+^{ab}=\frac{1}{\sqrt{2}}(|1\rangle_A|0\rangle_B+|0\rangle_A|1\rangle_B)\equiv \frac{1}{\sqrt{2}}(a^{\dagger}+b^{\dagger})|0\rangle,$ but due to phase errors, they share a state which has an admixture of the singlet state $\psi_-^{ab}$ 
\begin{equation}
\label{mixedstate}
 \rho_{ab}=F|\psi_+^{ab}\rangle\langle\psi_+^{ab}|+(1-F)|\psi_-^{ab}\rangle\langle\psi_-^{ab}|.
 \end{equation}
$F$ is the fidelity of the shared state : if $F=1/2,$ the phase information is lost and no entanglement is left while in the case where $F=1,$ the state is maximally entangled.

Note that for quantum repeaters, the phase errors are the most important. The empty component $|0\rangle_A|0\rangle_B$ does not affect the fidelity of the distributed state since the final step of single-photon protocols post-selects the cases where there was a photon in the output state \cite{Duan01}. The multi-photon components $|1\rangle_A|1\rangle_B$ can be greatly reduced using specific architectures \cite{Sangouard07}.

Suppose that Alice and Bob share two copies of the state described by (\ref{mixedstate}), $\rho_{a_1b_1}$ with fidelity $F_1$ and $\rho_{a_2b_2}$ with fidelity $F_2$ (see Fig. \ref{fig1}). Alice and Bob both perform unitary transformations on their modes $a_1,$ $a_2$ and $b_1,$ $b_2$ respectively: Alice combines the two modes $a_1,$ $a_2$ on a beam splitter with an intensity transmission of 85\% and Bob uses a beam splitter with an intensity transmission of 15\%. The detection of a single photon by Alice in mode $d_a$ (or by Bob in mode $d_b$), projects the modes $\tilde{a},$ $\tilde{b}$ on a mixed state $\rho_{\tilde{a}\tilde{b}}$ with fidelity 
\begin{equation}
\tilde{F}=\frac{F_1 F_2 + F_1/2 + F_2/2}{1 + F_1 F_2 + (1 - F_1) (1 - F_2)}.
\end{equation}
Remarkably, the state resulting of this simple operation is substantially purified. As an example, if errors are of the order of $\epsilon$, i.e. $F_1=F_2=1-\epsilon,$ the purification protocol divides them by a factor of 2, i.e. $\tilde{F}=1-\epsilon/2+o(\epsilon^2).$ In quantum repeaters the error is approximately doubled with every level of entanglement swapping. The present protocol has the potential to significantly increase the number of possible levels, and thus the achievable distance. Furthermore, in principle, the protocol could be applied again to the already purified states, and this process could continue as long as there are entangled states remaining, obtaining an increasingly purified state at every step. Note that the success probability for the purification protocol is $p=\frac{1}{4}(1 + F_1 F_2 + (1 - F_1) (1 - F_2))$ which is close to $1/2$ for $F_1$ and $F_2$ close to $1$. Note also that the previously mentioned proposal \cite{Pan01} based on polarization entanglement squares the errors, i.e. $\tilde{F}=1-\epsilon^2+o(\epsilon^3)$. We have shown that our scheme achieves the optimal fidelity (see Ref. \cite{Sangouard08pur} for a detailed discussion) for single-photon entanglement. This protocol reveals an intrinsic difference between single-photon entanglement and polarization-entangled photons. 

\begin{figure}
\includegraphics[width=1.0\linewidth]{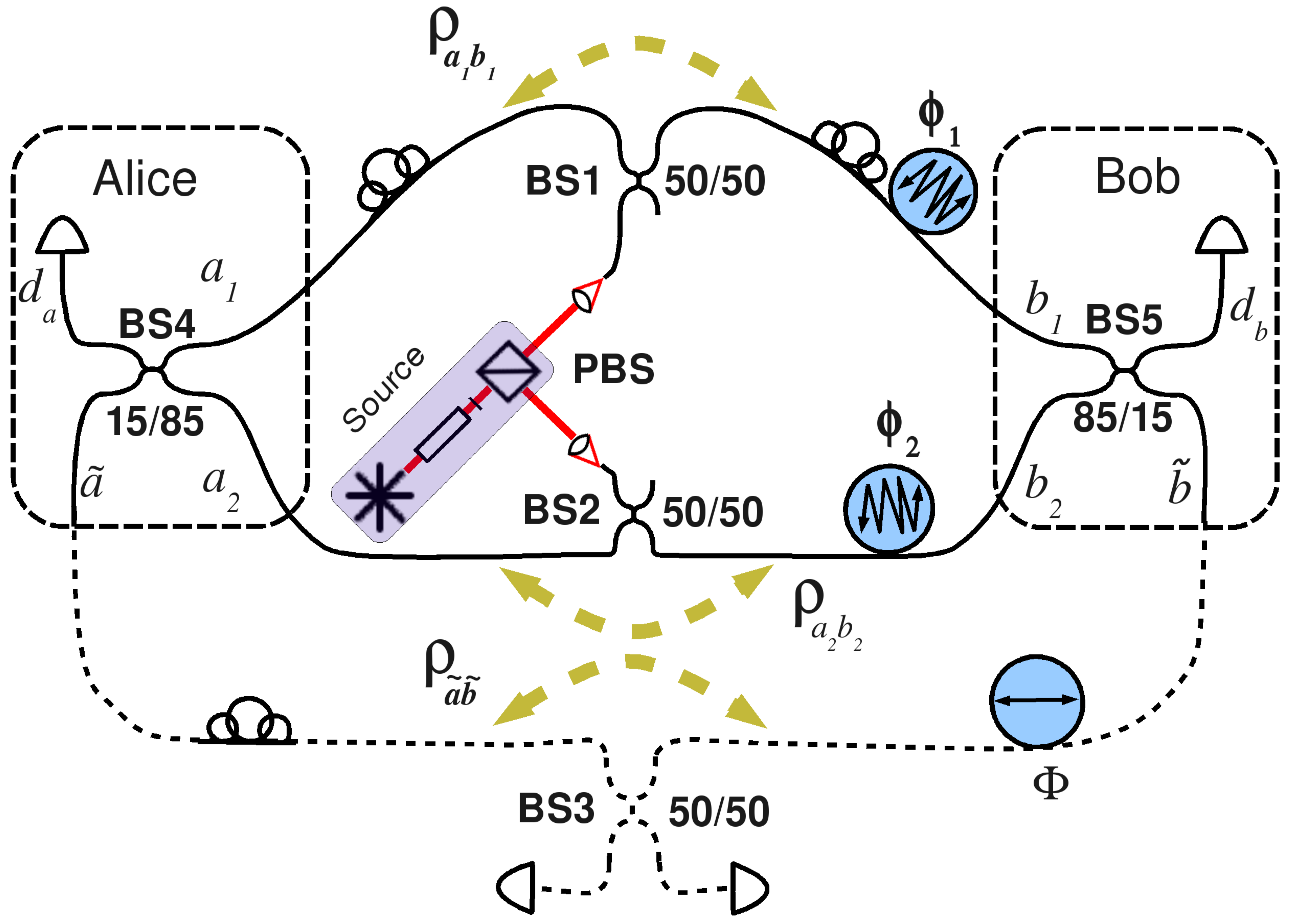}
\caption{\label{fig2} (color online). Experimental setup. Pairs of orthogonally polarized photons are created by a waveguide-based source, separated by a PBS and coupled into optical fibers. Each photon passes through a 50/50 coupler (BS1 and BS2) to create two-mode entangled states $\rho_{a_1b_1}$ and $\rho_{a_2b_2}$. Alice and Bob each receive two modes, one from each state, and combine them using 15/85 couplers (BS4 and BS5). Conditional on the detection of one photon by either Alice or Bob, a purified single-photon entangled state $\rho_{\tilde{a}\tilde{b}}$ is created. The degree of entanglement is measured using 50/50 coupler BS3. Two noise generators ($\phi_1$ and $\phi_2$) are used to reduce the fidelities $F_1$ and $F_2$ of $\rho_{a_1b_1}$ and $\rho_{a_2b_2}$, respectively. The phase $\Phi$ is scanned using a piezo to acquire an interferogram and thus estimate the fidelities.}
\end{figure}

The physics behind the purification is based on the interference of two modes sharing a single photon and on the bosonic character of indistinguishable photons. Single-photon interference requires a highly stable setup. Indistinguishability of the photons demands a good overlap of the temporal, spectral, spatial and polarization modes of the photons. Thus, the main challenge is the construction and operation of an experimental setup that allows us to obtain both indistinguishable photons and high visibility single-photon interference, leading for the first time to the purification of single-photon entanglement.

The experimental setup is shown in Fig. \ref{fig2}. A Ti-indiffused 7\,$\mu$m wide waveguide in 25\,mm long periodically poled ($\Lambda$ = 9.15\,$\mu$m) lithium niobate (Ti:PPLN) operated at 43°C - monomode around 1.5\,$\mu$m wavelength - creates degenerate photon pairs through the process of spontaneous parametric down conversion. The periodicity of $\Lambda$ has been chosen to get ``type-II" quasi phase matching (QPM) for orthogonally polarized signal and idler photons. The waveguide is pumped by a continuous-wave single mode external cavity diode laser at 780\,nm (Toptica DL100). After the waveguide, the remaining light from the pump laser is blocked by a silicon filter. The signal and idler photons, with a spectral width of 3\,nm (full-width at half-maximum) centered at 1560\,nm, both pass through the same narrowband filter with a bandwidth of 1.3\,nm reducing their spectral distinguishability. The photons are then separated with a polarizing beam splitter (PBS) and coupled into single mode optical fibers. Each photon is sent through a 50/50 coupler (BS1 and BS2) to prepare the two-mode entangled states $\rho_{a_1b_1}$ and $\rho_{a_2b_2}$. These states are then distributed between Alice and Bob. They each receive two modes, one from each entangled state, and combine them using couplers BS4 and BS5, respectively. These last two couplers are manual variable-ratio evanescent wave couplers (Canadian Instrumentation $\&$ Research Ltd). Finally, a 50/50 coupler (BS3) is used to measure the degree of entanglement (see below). The photons are detected by single-photon avalanche photodiodes (APD) and sent to an electronic coincidence counting circuit for analysis. The heralding detector is a free-running InGaAs/InP APD with home-made electronics \cite{Thew07}. We operate it at an efficiency of 8\%, 1300 dark counts\,$\cdot$\,s$^{-1}$ and a dead time of 30\,$\mu$s. It triggers an InGaAs APD (idQuantique, id200) at a rate of 15.5\,kHz working on gated mode and operated at an efficiency of 5.5\%, $2.7\times10^{-5}$ dark counts\,$\cdot$\,ns$^{-1}$ and a dead time of 10\,$\mu$s.

Using a pump laser power of $16$\,mW, we generate $1.3\times10^6$\,pairs\,$\cdot$\,s$^{-1}$ for a spectral width of 1.3\,nm, leading to a brightness of the source of $5\times10^2$ pairs\,$\cdot$\,s$^{-1}$\,$\cdot$\,GHz$^{-1}$\,$\cdot$\,mW$^{-1}$. The emission flux $p$ is $1.0\times10^{-3}$\,pairs per detection time $\tau_d=800$\,ps. The detection time is the FWHM of the coincidence peak and it depends primarily on the jitter of the single-photon detectors. 

The degree of indistinguishability of the two photons can be measured through the visibility of the ``Hong-Ou-Mandel dip" \cite{HOMdip}. If the photons were perfectly indistinguishable, the number of coincidences would be zero and the visibility of the HOM dip would be 100\%. We have estimated, using a simple model with discrete modes, that the presented protocol requires overlaps between the distributions associated to the modes $a_1,$ $a_2,$ $b_1$ and $b_2$ above 90\% to obtain a significant purification effect. The temporal overlap between the modes is achieved by adjusting the path lengths that each photon has to travel between the PBS and the couplers BS3, BS4 and BS5. The spectral overlap is ensured due to both photons passing through the same narrowband filter. The use of single-mode fibers guarantees the transverse spatial overlap. Lastly, the polarization is controlled using the polarization controllers shown in Fig. \ref{fig2}. After performing these adjustments, we observed a HOM dip with the extremely high visibility of $V_{dip}=(C_{max}-C_{min})/C_{max}=(99.0\pm0.3)\%$.

To determine the degree of entanglement for $\rho_{a_1b_1}$, the reflected photons at the PBS are not sent to BS2 but detected to herald the creation of the state $\rho_{a_1b_1}$ at coupler BS1. The modes $a_1$, $b_1$ are combined at coupler BS3. The path chosen by the single photon is unknown, leading to interference fringes when the phase $\Phi$ of the interferometer is scanned. The visibility $V_1$ of the interference fringes gives a good estimation of the fidelity $F_1=(1+V_1)/2$ of the state $\rho_{a_1b_1}$ since we postselect the cases where there is at least one excitation in either $a_1$ or $b_1$ and since the multi-pair emissions are weakly probable, as confirmed by the visibility of the HOM dip. To scan this phase, we use a circular piezoelectric actuator with optical fiber coiled around it. A voltage ramp is applied to the piezo that progressively expands, stretching the fiber and changing the phase. To determine the degree of entanglement for $\rho_{a_2b_2}$, the measurement is repeated inverting the roles for the transmitted and reflected photons.

\begin{figure}
\includegraphics[width=1.0\linewidth]{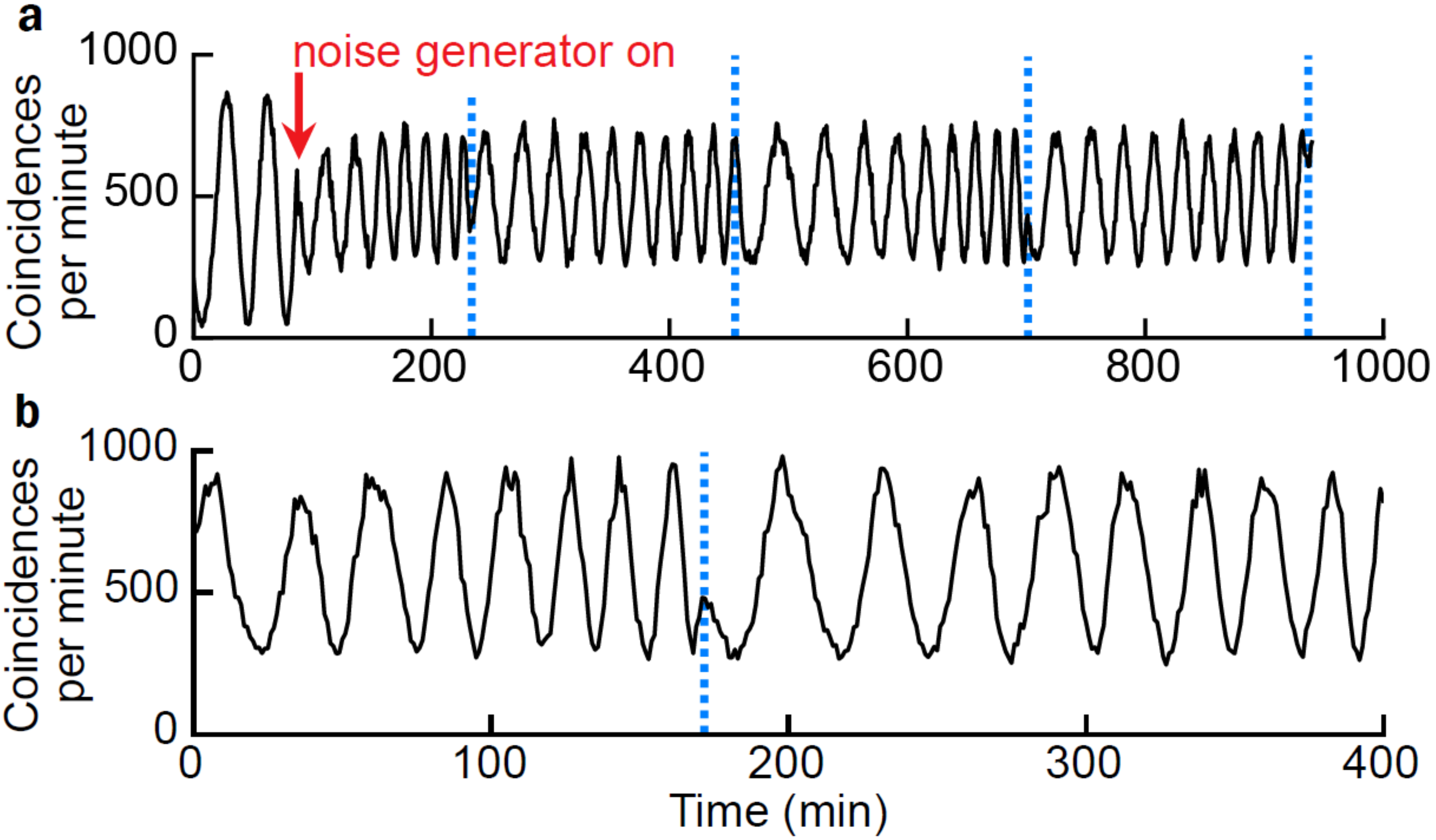}
\caption{\label{fig3} (color online). Interference fringes observed while the phase $\Phi$ is being scanned. (a) For the state $\rho_{a_2b_2}$, initially, the fidelity is $F_2=(97.7\pm0.2)$\%. When the noise generator $\phi_2$ is switched on, the fidelity decreases to $F_2=(75.0\pm0.7)$\%. (b) For the purified state $\rho_{\tilde{a}\tilde{b}}$, while both noise generators are on, the fidelity is $\tilde{F}=(79.6\pm1.1)$\%. The vertical lines mark every time the voltage ramp reaches its end, momentarily interrupting the scan, and reversing the scan direction.}
\end{figure} 

Note that there is no need for a phase reference. We only need to avoid uncontrolled phase oscillations to be able to perform the purification. The length of the interferometric path from the PBS to the BS3 is $10\,m$. The phase stability on this distance was maintained just by keeping the temperature of the setup stable. For longer distances, an active stabilization system, such as the one reported in \cite{ChoandNoh} for a 6-km long interferometer, would probably be required to keep the phase of the interferometer stable.

To simulate field conditions in a controlled and reproducible way, noise is introduced in the interferometers. The purification protocol that we are testing works for a wide range of fidelities $F_1$ and $F_2$, but we reduce these fidelities closer to $F_1=F_2=76\%$ where the fidelity increase is greatest \cite{Sangouard08pur}. The function used to introduce the noise was chosen to reproduce the gaussian phase-noise distribution in a fiber, as observed in real world networks \cite{Minar07}. To generate this noise, we use two additional circular piezos ($\phi_1$ for state $\rho_{a_1b_1}$ and $\phi_2$ for state $\rho_{a_2b_2}$) that vibrate at a frequency much higher than the integration time of the measurement. This noise is independently generated for each piezo. Interference fringes measured in one of the entangled states and the reduction of the fidelity due to the noise generation are shown in Fig. \ref{fig3}a.

To prepare the purified state, the variable couplers BS4 and BS5 are adjusted to the intensity transmissions required to apply the purification protocol \cite{Sangouard08pur}, corresponding to 85\% for Alice and 15\% for Bob (or vice versa). Modes $a_1$ and $a_2,$ are combined by Alice to form modes $\tilde{a}$ and $d_a$, while modes $b_1$ and $b_2$ are combined by Bob to form modes $\tilde{b}$ and $d_b$. Conditioned on the detection of one photon on either $d_a$ or $d_b$, the modes $\tilde{a}$ and $\tilde{b}$ are combined at coupler BS3. Again, because we can not know which path the photons have taken, interference fringes are observed when the phase $\Phi$ is scanned (see Fig. \ref{fig3}b). As before, the fidelity $\tilde{F}$ of the state $\rho_{\tilde{a}\tilde{b}}$ is deduced from the visibility of the fringes.

\begin{figure}
\includegraphics[width=1.0\linewidth]{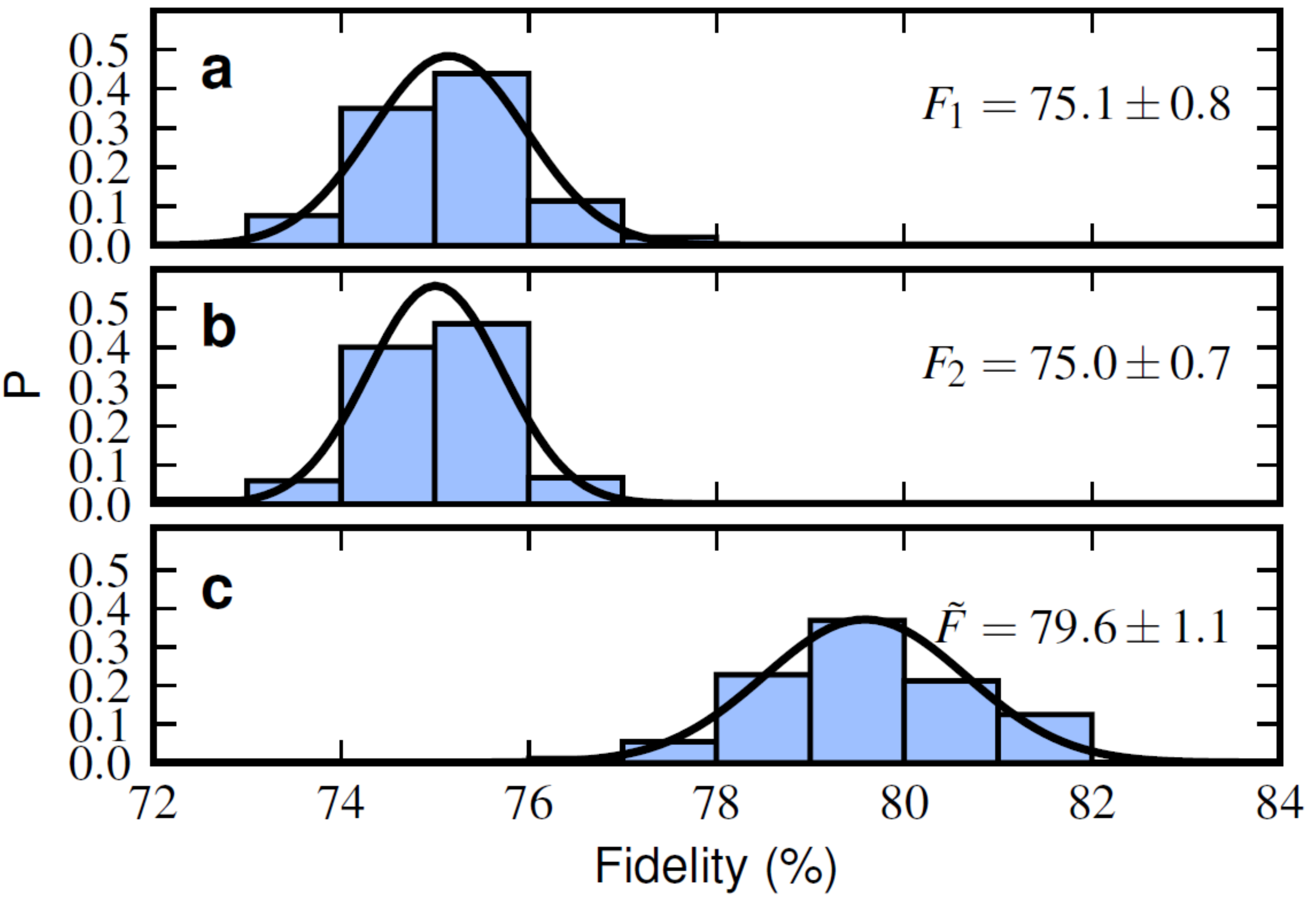}
\caption{\label{fig4} (color online). Distribution of the fidelity measurements. Probability densities P as a function of the measured fidelities for (a) state $\rho_{a_1b_1}$, (b) state $\rho_{a_2b_2}$ and (c) the purified state $\rho_{\tilde{a}\tilde{b}}$. The corresponding mean fidelity and standard deviation are given next to each distribution. The curves are gaussian functions with the same mean and $\sigma$ as the sample data.}
\end{figure}

For each of the entangled states ($\rho_{a_1b_1}$, $\rho_{a_2b_2}$ and $\rho_{\tilde{a}\tilde{b}}$), measurements of several interference fringes  were obtained. Using sequential sinusoidal fits of approximately two periods, we calculated the fidelities for all fringes. The resulting distributions of fidelity values are represented in Fig. \ref{fig4}. From each set of values, the mean fidelities $F_1$, $F_2$ and $\tilde{F}$ were calculated. The given uncertainty values are the standard deviations ($\sigma$) associated with each distribution. More details on the measurement and the data analysis are provided in the Supplemental Information. 

The initial fidelity of $\rho_{a_1b_1}$ is found to be $F_1=(97.8\pm0.2)$\% while $\rho_{a_2b_2}$ has a fidelity $F_2=(97.7\pm0.2)$\%. The residual 2\% are mainly due to path length instabilities. By introducing a controlled amount of noise that reproduces the phase-noise in a fiber, the fidelities reduce to $F_1=(75.1\pm0.8)$\% and $F_2=(75.0\pm0.7)$\%. By implementing the purification protocol, we distill a state $\rho_{\tilde{a}\tilde{b}}$ with fidelity $\tilde{F}=(79.6\pm1.1)$\%. The improvement in the degree of entanglement, taken as the difference between $\tilde{F}$ and $F_1$, is as high as 4.5\%. Note that it has been shown in Ref. \cite{Sangouard08pur} that the optimal theoretical value is of 5.7\%. The remaining 1.2\% is mainly due to imperfections in the mode overlap (see the Supplemental Information for details). Note that these values are obtained after subtracting accidental coincidences due to dark counts. As shown in Fig. \ref{fig4}, the overlap between the distributions of initial and purified fidelity values is negligible, leaving no doubt about the influence of the purification effect. Furthermore, this effect remains clear even when the accidental coincidences are not subtracted. See the Supplemental Information for detailed discussions of the imperfect fidelities and the accidental coincidences.

Single-photon entanglement has been at the heart of a lively debate \cite{Tan,vanEnk}. Part of the controversy has been solved by mapping single-photon entanglement into two atomic ensembles and by revealing the entanglement between these ensembles \cite{Choi}. Note also that entanglement between four modes sharing a single photon has been characterized by direct measurements of the optical modes \cite{Papp}. Our experiment further shows that single-photon entanglement can be purified using linear optics. Looking further ahead, this simple protocol might also become useful in the frame of quantum repeaters. In such an experiment, single-photon entangled states would be used to create entanglement between two pairs of atomic ensembles.  Then, the stored excitations would be reconverted into photons, combined using linear optics and detected. Additional experimental challenges would be to reabsorb the purified delocalized photon in a quantum memory and, of course, to increase the distance between the two parties.\\

{\large \bf Supplemental Information}\\

\noindent
\textbf{Measurement procedure.}\\
The first step of the measurement procedure was to start the scan of the interferometer phase with the noise generator switched off. The first few interference fringes were used to confirm that the fidelity remained high. Without interruption, the noise generator was switched on and we recorded interference fringes with a reduced fidelity. The number of consecutive fringes in a measurement depended on the thermal stability of the setup. We measured up to 33 consecutive fringes for state $\rho_{a_1b_1}$, 34 for state $\rho_{a_2b_2}$ and 17 for state $\rho_{\tilde{a}\tilde{b}}$. These are the measurements used for the reported data in this Letter. We also observed that the period of fringes becomes longer every time that the phase scan direction is reversed (see Fig. 3 in the article). This is attributed to an inertia effect from either the scanning piezo or the fiber coiled around it. We made sure that the measured values were repeatable (within the margin of error) and that the initial fidelities could be recovered after the noise generator was switched back off at the end of a measurement.\\

\noindent
\textbf{Data analysis.}\\
The analysis of the measured data was performed as follows. The fringes were fitted with a sinusoidal function with a length of approximately two periods that was sequentially displaced across the measurement. From the large number of fits obtained, we made sure that the fits used for the data analysis followed the measured points well. This meant discarding a minority of fits corresponding to points where the function was unable to fit the fringes correctly, for example, every time that the phase scan was interrupted because the voltage ramp had reached its end (see Fig. 3 in the article). Each fit gave us a visibility value from which the fidelity was calculated. To compare the degree of entanglement between different measurements, we calculated the mean fidelity. We took the standard deviation of the fidelities as the uncertainty of each measurement. If the spread of the fidelity values is not the product of a systematic error in the measurement, then the distribution of fidelities should take a gaussian shape. As shown in Fig. 4 in the article, there is a good agreement between the distribution of fidelity values and the gaussian curves. 

The uncertainties for $F_1$ and $F_2$ obtained when the noise generator is switched on are larger than the ones obtained when it is switched off. This is explained by considering that the uncertainty depends on the ratio of the amplitude error vs the amplitude of the fringes. This ratio takes a larger value when the amplitude is smaller (as is the case when the noise generator is switched on), leading to larger uncertainties.\\

\noindent
\textbf{Accidental coincidences.}\\
Accidental coincidences occur when we get a coincidental count that is not the result of two photons coming from the source. Most of the times, it is a dark count at one of the detectors and a single-photon at the other, with the cases where it is one dark count at each detector being only a minority. We measured 26.6 accidental coincidences per minute for the measurement on state $\rho_{a_1b_1}$, 27.2 on state $\rho_{a_2b_2}$ and 32.3 on state $\rho_{\tilde{a}\tilde{b}}$. They were subtracted from the number of total coincidences before the analysis of the measured data. If accidentals are not subtracted, the measured fidelities are $F_1=(73.9\pm0.8)$\%, $F_2=(73.7\pm0.7)$\% and $\tilde{F}=(77.8\pm1.1)$\%, corresponding to an increase of 3.9\% in the purified state fidelity.\\

\noindent
\textbf{Imperfect HOM dip visibility and fidelities.}\\
The reduction in the initial fidelities $F_1$ and $F_2$ of the order of 2\% is mainly due to fluctuations in the temporal overlap of modes $\tilde{a}$ and $\tilde{b}$ involved in the single photon interferences, which are interferometrically sensitive to phase fluctuations. This is confirmed by the measurement of values for $F_1$ and $F_2$ above 99\% without heralding. In this case, the count rates are higher, leading to smaller integration times (and thus smaller phase fluctuations). Note that imperfections in the overlap of the polarization modes can also reduce the fidelities $F_1$ and $F_2$.

The visibility of the HOM dip has been measured at both Alice's and Bob's locations. On the basis of a simple model with discrete time units of duration $\tau_d$, the reduction in visibility of the HOM dip due to double-pair emission is estimated to be of the order of $2p/(1+3p)\sim0.2$\%. This means that, as the purification requires, the modes overlap very well, accounting for a reduction of the dip visibility of $\sim0.8$\% only. However, this measurement does not guarantee that the phase of modes involved in the HOM dip is interferometrically stable.

The reduction in the purified fidelity $\tilde{F}$ of the order of 1.2\% is primarily due to the imperfection in the mode overlap, which accounts at least for a reduction of 0.25\%. We underestimated this effect because we only considered the partial overlap between $a_1$ and $a_2$, $b_1$ and $b_2$, but we did not take into account the imperfections in the overlap between $\tilde{a}$ and $\tilde{b}$. We also estimate that double-pair emission accounts for a reduction of the order of 0.05\%. The noise of the detector heralding the purification effect leads to an additional reduction in the fidelity of the order of 0.05\%. Concerning the remaining error (of the order of 0.8\%), we believe that it is caused by phase fluctuations of modes involved in the single photon interferences due to long integration times. Note also that due to the uncertainty in the transmission of the variable couplers, there is a systematic error of 0.6\% on the estimate of the purified fidelity.\\

We thank C. Barreiro and J.-D. Gautier for technical support and H. de Riedmatten for his insightful comments and careful reading of the manuscript. This work was supported by the Swiss NCCR Quantum Photonics and the European Union projects QAP and ERC-AG QORE.

\end{document}